\documentclass[manuscript]{aastex}

\newcommand{\myemail}{abe@stelab.nagoya-u.ac.jp}

\shorttitle{Gravitational Microlensing by the Ellis Wormhole}
\shortauthors{F. Abe}

\begin{document}


\title{Gravitational Microlensing by the Ellis Wormhole}


\author{F. Abe\altaffilmark{1}}
\affil{Solar-Terrestrial Environment Laboratory, Nagoya University \\
Furo-cho, Chikusa-ku, Nagoya 464-8601, Japan \\
\myemail}


\altaffiltext{1}{Nagoya University Southern Observatories}


\begin{abstract}
A method to calculate light curves of the gravitational microlensing of the Ellis wormhole is derived in the weak-field limit. In this limit, lensing by the wormhole produces one image outside the Einstein ring and one other image inside. The weak-field hypothesis is a good approximation in Galactic lensing if the throat radius is less than $10^{11} km$. The light curves calculated have gutters of approximately 4\% immediately outside the Einstein ring crossing times. The magnification of the Ellis wormhole lensing is generally less than that of Schwarzschild lensing. The optical depths and event rates are calculated for the Galactic bulge and Large Magellanic Cloud fields according to bound and unbound hypotheses. If the wormholes have throat radii between $100$ and $10^7 km$, are bound to the galaxy, and have a number density that is approximately that of ordinary stars, detection can be achieved by reanalyzing past data. If the wormholes are unbound, detection using past data is impossible.
\end{abstract}




\section{Introduction}
A solution of the Einstein equation that connects distant points of space--time was introduced by \citet{ein35}. This "Einstein--Rosen bridge" was the first solution to later be referred to as a wormhole. Initially, this type of solution was just a trivial or teaching example of mathematical physics. However, \citet{morr88} proved that some wormholes are "traversable"; i.e., space and time travel can be achieved by passing through the wormholes. They also showed that the existence of a wormhole requires exotic matter that violates the null energy condition. Although they are very exotic, the existence of wormholes has not been ruled out in theory. Inspired by the Morris--Thorne paper, there have been a number of theoretical works (see \citet{vis95, lob09} and references therein) on wormholes. The curious natures of wormholes, such as time travel, energy conditions, space--time foams, and growth of a wormhole in an accelerating universe have been studied. Although there have been enthusiastic theoretical studies, studies searching for real evidence of the existence of wormholes are scarce. Only a few attempts have been made to show the existence or nonexistence of wormholes.

A possible observational method that has been proposed to detect or exclude the existence of wormholes is the application of optical gravitational lensing. The gravitational lensing of wormholes was pioneered by \citet{cram95}, who inferred that some wormholes show "negative mass" lensing. They showed that the light curve of the negative-mass lensing event of a distant star has singular double peaks. Several authors subsequently conducted theoretical studies on detectability \citep{saf01, bog08}. Another gravitational lensing method employing gamma rays was proposed by \citet{tor98}, who postulated that the singular negative-mass lensing of distant active galactic nuclei causes a sharp spike of gamma rays and may be observed as double-peaked gamma-ray bursts. They analyzed BASTE data and set a limit for the density of the negative-mass objects.

 There have been several recent works \citep{sh04, per04, nan06, rah07, dey08} on the gravitational lensing of wormholes as structures of space--time. Such studies are expected to unveil lensing properties directly from the space--time structure. One study \citet{dey08} calculated the deflection angle of light due to the Ellis wormhole, whose asymptotic mass at infinity is zero. The massless wormhole is particularly interesting because it is expected to have unique gravitational lensing effects.
The Ellis wormhole is expressed by the line element
\begin{equation}
ds^2 = dt^2 - dr^2 - (r^2 + a^2) (d\theta^2 + sin^2(\theta) d\phi^2),
\end{equation}
where $a$ is the throat radius of the wormhole. This type of wormhole was first introduced by \cite{ell73} as a massless scalar field. 
Later, \citet{morr88} studied this wormhole and proved it to be traversable. The dynamical feature was studied by \cite{shi02}, who showed that Gaussian perturbation causes either explode to an inflationary universe or collapse to a black hole. \cite{das05} showed that the tchyon condensate can be a source for the Ellis geometry.

In this paper, we derive the light curve of lensing by the Ellis wormhole and discuss its detectability. In Section 2, we discuss gravitational lensing by the Ellis wormhole in the weak-field limit. The light curves of wormhole events are discussed in Section 3. The validity of the weak-field limit is discussed in Section 4.  The optical depth and event rate are discussed in Section 5. The results are summarized in Section 6.

\section{Gravitational lensing}
Magnification of the apparent brightness of a distant star by the gravitational lensing effect of another star was predicted by \citet{ein36}. This kind of lensing effect is called "microlensing" because the images produced by the gravitational lensing are very close to each other and are difficult for the observer to resolve. The observable effect is the changing apparent brightness of the source star only. This effect was discovered in 1993 \citep{uda93, alc93, aub93} and has been used to detect astronomical objects that do not emit observable signals (such as visible light, radio waves, and X rays) or are too faint to observe. Microlensing has successfully been applied to detect extrasolar planets \citep{bon04} and brown dwarfs \citep{nov09, gou09}. Microlensing is also used to search for unseen black holes \citep{alc01, ben02, poi05} and massive compact halo objects \citep{alc00, tis07, wyr09}, a candidate for dark matter.

\begin{figure}
\epsscale{.90}
\plotone{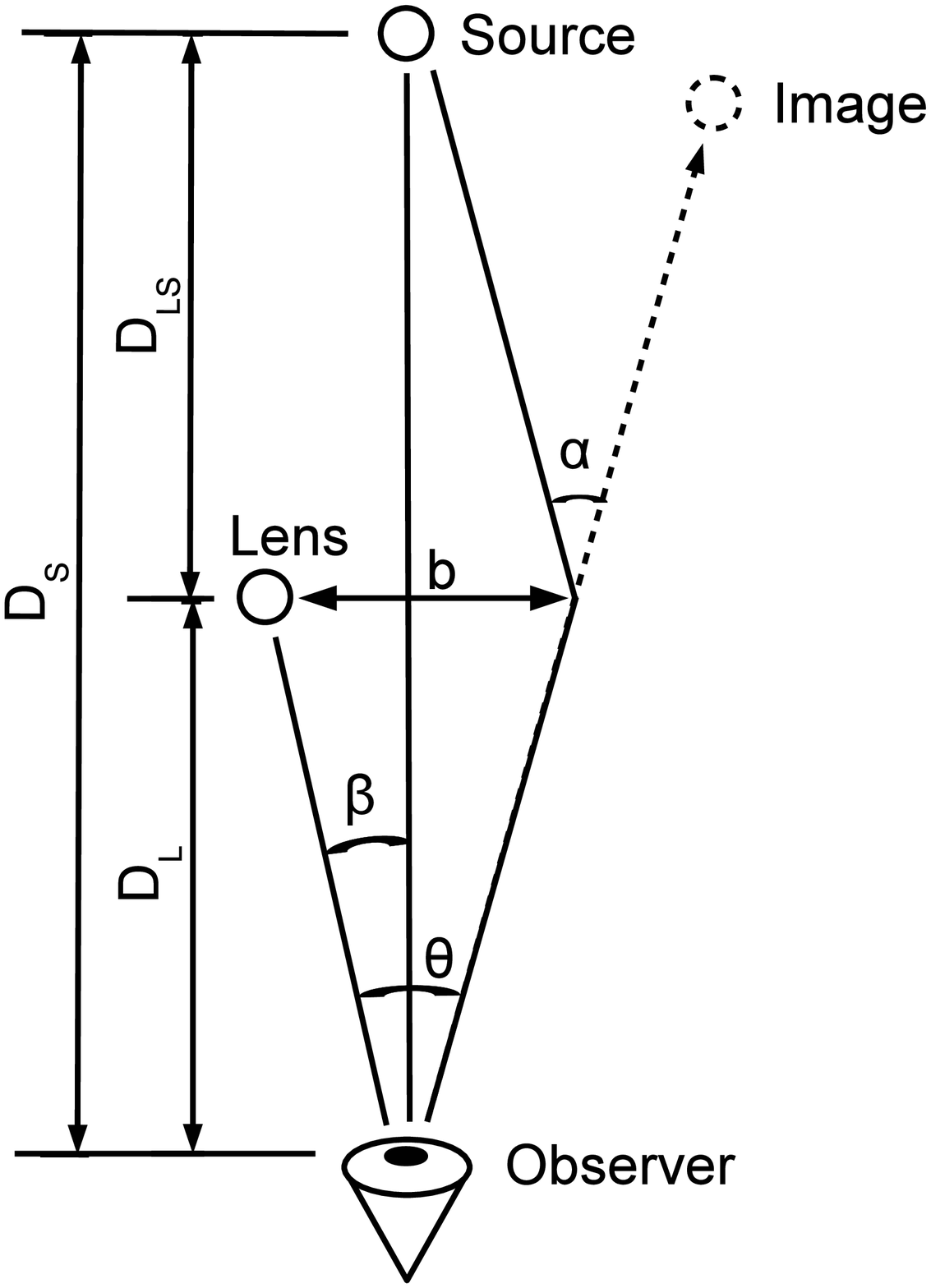}
\caption{Sketch of the relation between the source star, lens (wormhole), and observer. \label{fig1}}
\end{figure}

The gravity of a star is well expressed by the Schwarzschild metric. The gravitational microlensing of the Schwarzschild metric \citep{ref64, lieb64, Pacz86} has been studied in the weak-field limit. In this section, we simply follow the method used for Schwarzschild lensing. Figure \ref{fig1} shows the relation between the source star, the lens (wormhole), and the observer. The Ellis wormhole is known to be a massless wormhole, which means that the asymptotic mass at infinity is zero. However, this wormhole deflects light by gravitational lensing \citep{cle84,che84,nan06,dey08} because of its curved space--time structure. The deflection angle $\alpha(r)$ of the Ellis wormhole was derived by \cite{dey08} to be
\begin{equation}
\alpha(r) = \pi \left\{\sqrt \frac{2 (r^2 + a^2)}{2 r^2 + a^2} -1 \right\},
\end{equation}
where $r$ is the closest approach of the light.
In the weak-field limit ($r \rightarrow \infty$), the deflection angle becomes
\begin{equation}
\alpha(r) \rightarrow \frac{\pi}{4}\frac{a^2}{r^2} - \frac{5 \pi}{32} \frac{a^4}{r^4} + o\left(\frac{a}{r}\right)^6. \label{eqn:defl}
\end{equation}

The angle between the lens (wormhole) and the source $\beta$ can then be written as
\begin{equation}
\beta  = \frac{1}{D_L} b - \frac{D_{LS}}{D_S}\alpha(r),
\end{equation}
where $D_L$, $D_S$, $D_{LS}$, and $b$ are the distances from the observer to the lens, from the observer to the source, and from the lens to the source, and the impact parameter of the light, respectively. In the asymptotic limit, Schwarzschild lensing and massive Janis--Newman--Winnicour (JNW) wormhole lensing \citep{dey08} have the same leading term of $o\left(1/r\right)$. Therefore, the lensing property of the JNW wormhole is approximately the same as that of Schwarzschild lensing and is difficult to distinguish. As shown in Equation (\ref{eqn:defl}), the deflection angle of the Ellis wormhole does not have the term of $o\left(1/r\right)$ and starts from $o\left(1/r^2\right)$. This is due to the massless nature of the Ellis wormhole and indicates the possibility of observational discrimination from the ordinary gravitational lensing effect. In the weak-field limit, $b$ is approximately equal to the closest approach $r$. For the Ellis wormhole, $b = \sqrt{r^2 + a^2} \rightarrow r (r \rightarrow \infty)$. We thus obtain
\begin{equation}
\beta = \frac{r}{D_L} - \frac{\pi}{4} \frac{D_{LS}}{D_S}\frac{a^2}{r^2}\hspace{1cm}(r > 0). \label{eqn:proj} 
\end{equation}
 The light passing through the other side of the lens may also form images. However, Equation (\ref{eqn:proj}) represents deflection in the wrong direction at $r < 0$. Thus, we must change the sign of the deflection angle: 
\begin{equation}
\beta = \frac{r}{D_L} + \frac{\pi}{4} \frac{D_{LS}}{D_S}\frac{a^2}{r^2}\hspace{1cm}(r < 0). 
\label{eqn:proj2} 
\end{equation}
It would be useful to note that a single equation is suitable both for $r > 0$ and $r < 0$ images in the Schwarzschild lensing. However, such treatment is applicable only when the deflection angle is an odd function of $r$. 
 
 If the source and lens are completely aligned along the line of sight, the image is expected to be circular (an Einstein ring). The Einstein radius $R_E$, which is defined as the radius of the circular image on the lens plane, is obtained from Equation (\ref{eqn:proj}) with $\beta = 0$ as
 \begin{equation}
 R_E = \sqrt[3]{\frac{\pi}{4}\frac{D_L D_{LS}}{D_S} a^2}. \label{eqn:re}
 \end{equation}
 
The image positions can then be calculated from
\begin{equation}
\beta = \theta - \frac{\theta_E^3}{\theta^2} \hspace{1cm} (\theta > 0)  \label{eqn:imgeq}
\end{equation}
and 
\begin{equation}
\beta = \theta + \frac{\theta_E^3}{\theta^2} \hspace{1cm} (\theta < 0),  \label{eqn:imgeq2}
\end{equation}
where $\theta = b / D_L  \thickapprox r / D_L$ is the angle between the image and lens, and $\theta_E = R_E / D_L$ is the angular Einstein radius. Using reduced parameters $\hat{\beta} = \beta / \theta_E$ and $\hat{\theta} = \theta / \theta_E$, Equations (\ref{eqn:imgeq}) and (\ref{eqn:imgeq2}) become simple cubic formulas: 
\begin{equation}
\hat{\theta}^3 - \hat{\beta} \hat{\theta}^2 -1 = 0 \hspace{1cm} (\hat{\theta} > 0) \label{eqn:poly}
\end{equation}
and
\begin{equation}
\hat{\theta}^3 - \hat{\beta} \hat{\theta}^2 +1 = 0 \hspace{1cm} (\hat{\theta} < 0). \label{eqn:poly2}
\end{equation}
 As the discriminant of Equation (\ref{eqn:poly}) is $-4\hat{\beta}^3 - 27 < 0$, Equation (\ref{eqn:poly}) has two conjugate complex solutions and a real solution:
 \begin{equation}
 \hat{\theta} = \frac{\hat{\beta}}{3} + U_{1+} + U_{1-} ,
 \end{equation} 
 with, 
 \begin{equation}
 U_{1\pm} = \sqrt[3]{\frac{\hat{\beta}^3}{27} + \frac{1}{2} \pm \sqrt{\frac{1}{4}\left(1 + \frac{2 \hat{\beta}^3}{27}\right)^2 - \frac{\hat{\beta}^6}{27^2}}}.
 \end{equation}
The real positive solution corresponds to the physical image. 

The discriminant of Equation (\ref{eqn:poly2}) is $4\hat{\beta}^3 - 27$. Thus it has a real solution if $\hat{\beta} < \sqrt[3]{27/4}$: 
\begin{equation}
\hat{\theta} = \frac{\hat{\beta}}{3} + U_{2+} + U_{2-}, 
\end{equation}
where, 
\begin{equation}
U_{2\pm} = \omega \sqrt[3]{\frac{\hat{\beta}^3}{27} - \frac{1}{2} \pm \sqrt{\frac{1}{4}\left(1 - \frac{2 \hat{\beta}^3}{27}\right)^2 - \frac{\hat{\beta}^6}{27^2}}} , 
\end{equation}
with $\omega = e^{(2\pi/3)i}$. This solution corresponds to a physical image inside the Einstein ring. For $\hat{\beta} > \sqrt[3]{27/4}$, Equation (\ref{eqn:poly2}) has three real solutions. However, two of them are not physical because they do not satisfy $\hat{\theta} < 0$. Only the solution 
\begin{equation}
\hat{\theta} = \frac{\hat{\beta}}{3} +  \omega U_{2+} + U_{2-}
\end{equation}
corresponds to a physical image inside the Einstein ring. 

\begin{figure}
\includegraphics[angle=0,scale=1.0]{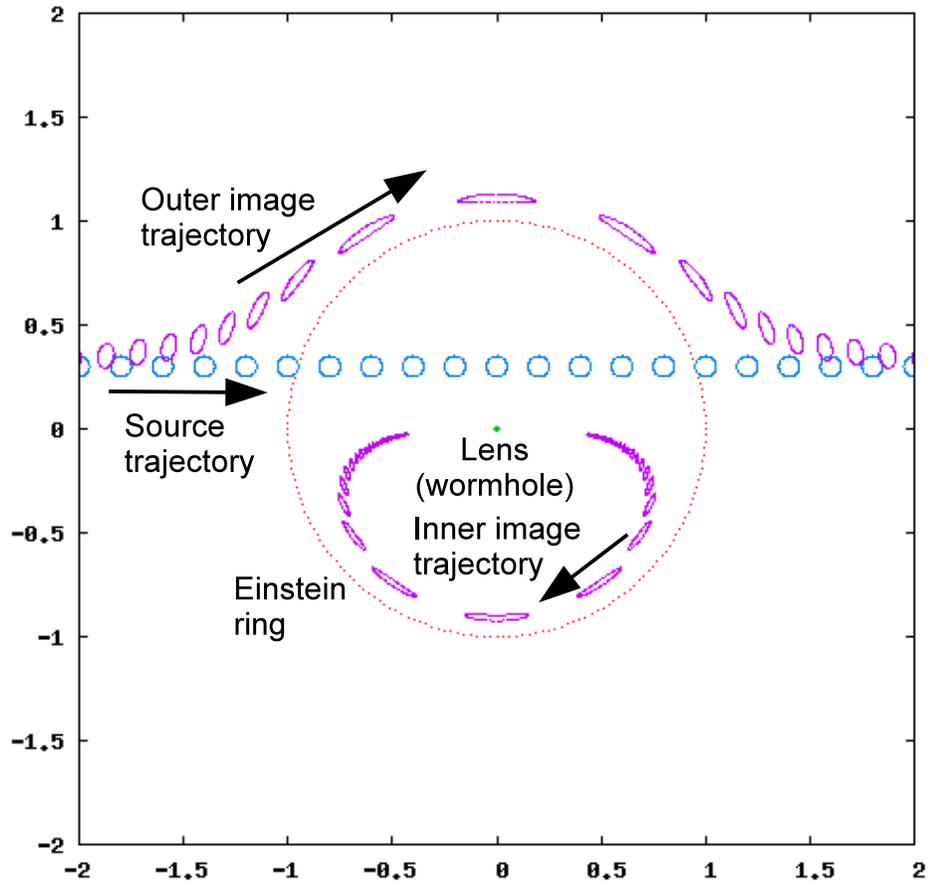}
\caption{Source and image trajectories in the sky from the position of the observer.\label{fig2}}
\end{figure}

 Figure \ref{fig2} shows the calculated images for source stars at various positions on a straight line (source trajectory). The motion of the images are similar to those of the Schwarzschild lensing. Table \ref{tbl-1} shows the Einstein radii and angular Einstein radii for a bulge star ($D_S = 8 kpc$ and $D_L = 4 kpc$ are assumed) and a star in the Large Magellanic Cloud (LMC, $D_S = 50 kps$ and $D_L = 25 kpc$ are assumed) for various throat radii. The detection of a lens for which the Einstein radius is smaller than the star radius ($\thickapprox 10^6 km$) is very difficult because most of the features of the gravitational lensing are smeared out by the finite-source effect. Thus, detecting a wormhole with a throat radius less than $1 km$ from the Galactic gravitational lensing of a star is very difficult. 

\section{Light curves}
\begin{figure}
\includegraphics[angle=0,scale=0.8]{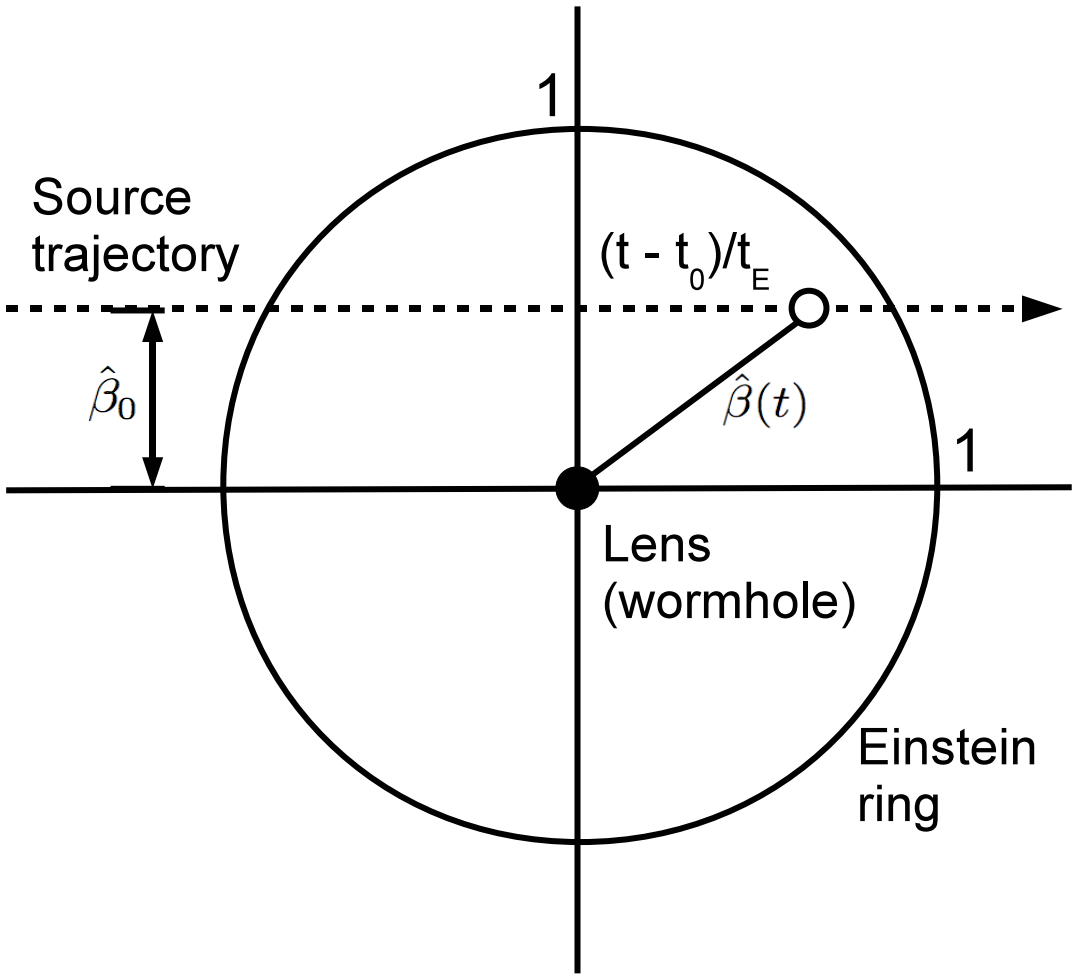}
\caption{Sketch of the relation between the source trajectory and the lens (wormhole) in the sky. All quantities are normalized by the angular Einstein radius $\theta_E$.\label{fig3}}
\end{figure}

\begin{figure}
\includegraphics[angle=0,scale=0.5]{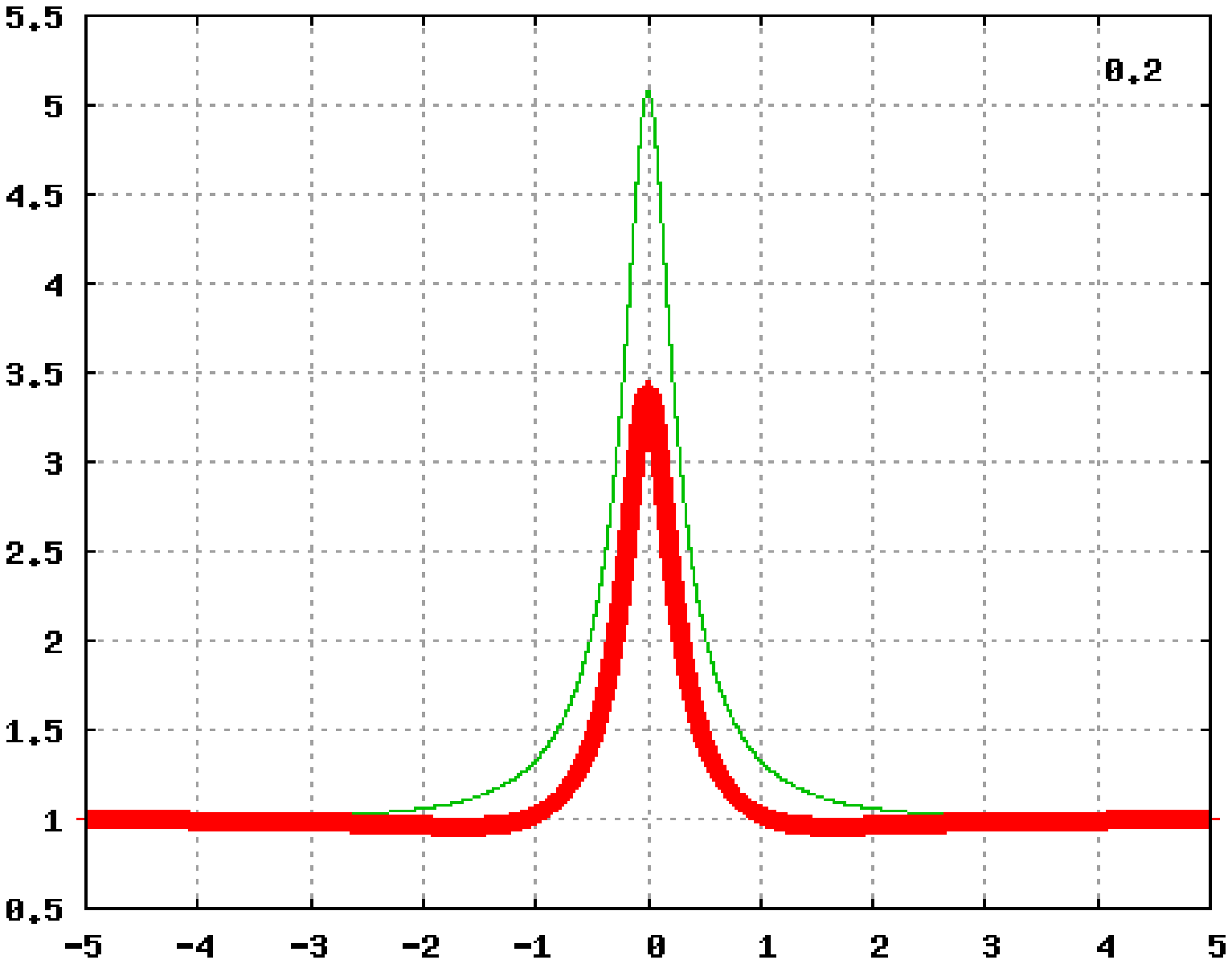}
\includegraphics[angle=0,scale=0.5]{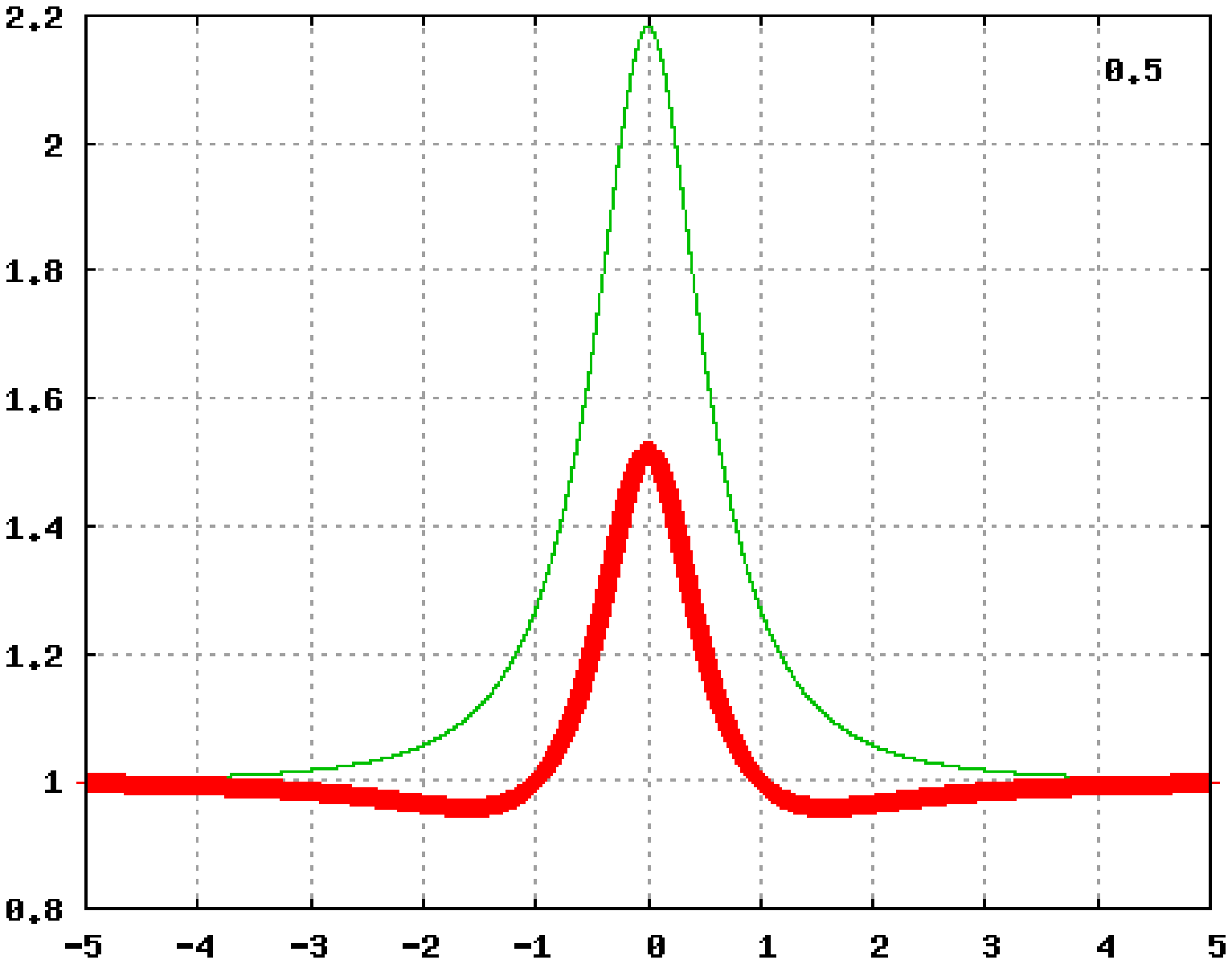}
\includegraphics[angle=0,scale=0.5]{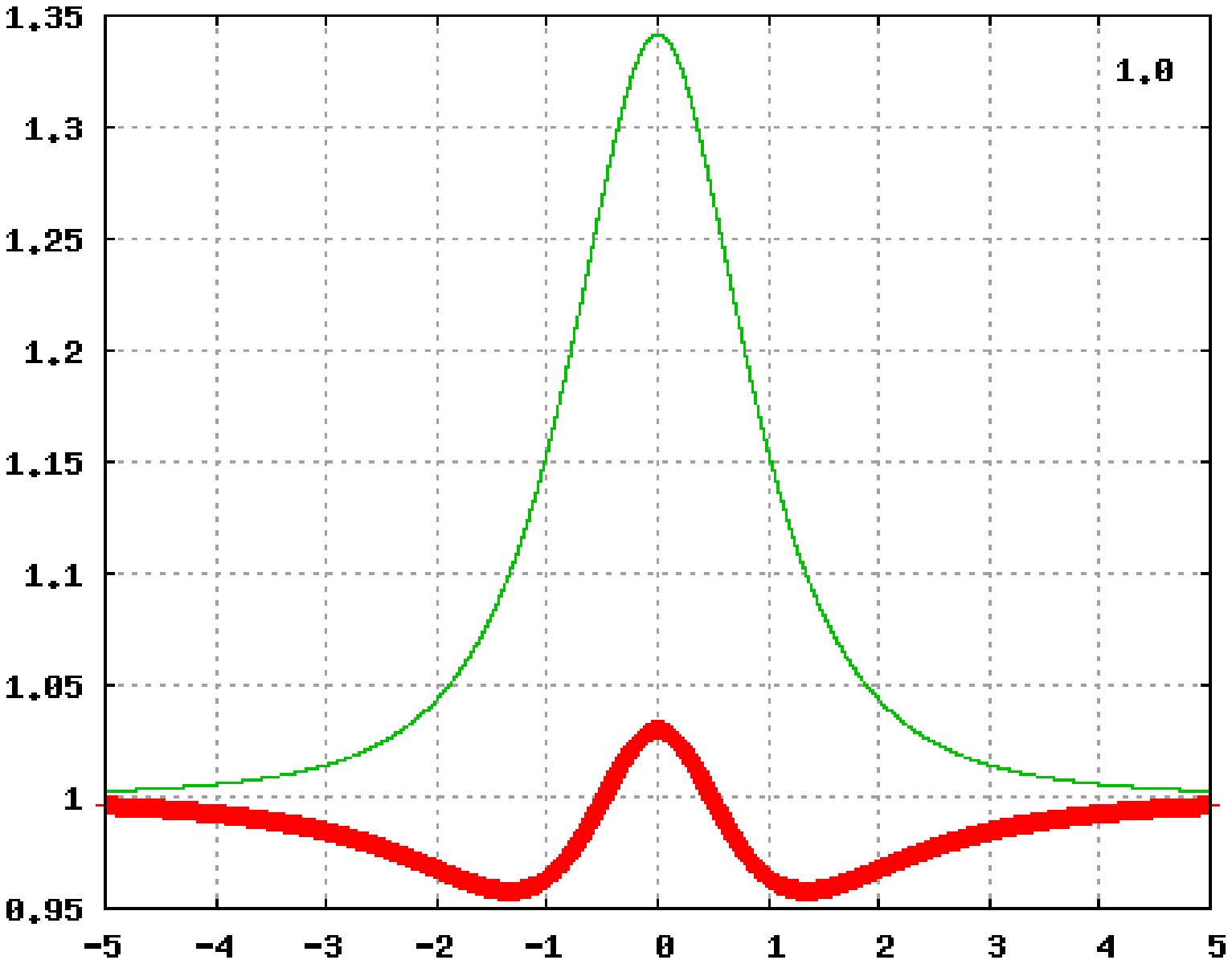}
\includegraphics[angle=0,scale=0.5]{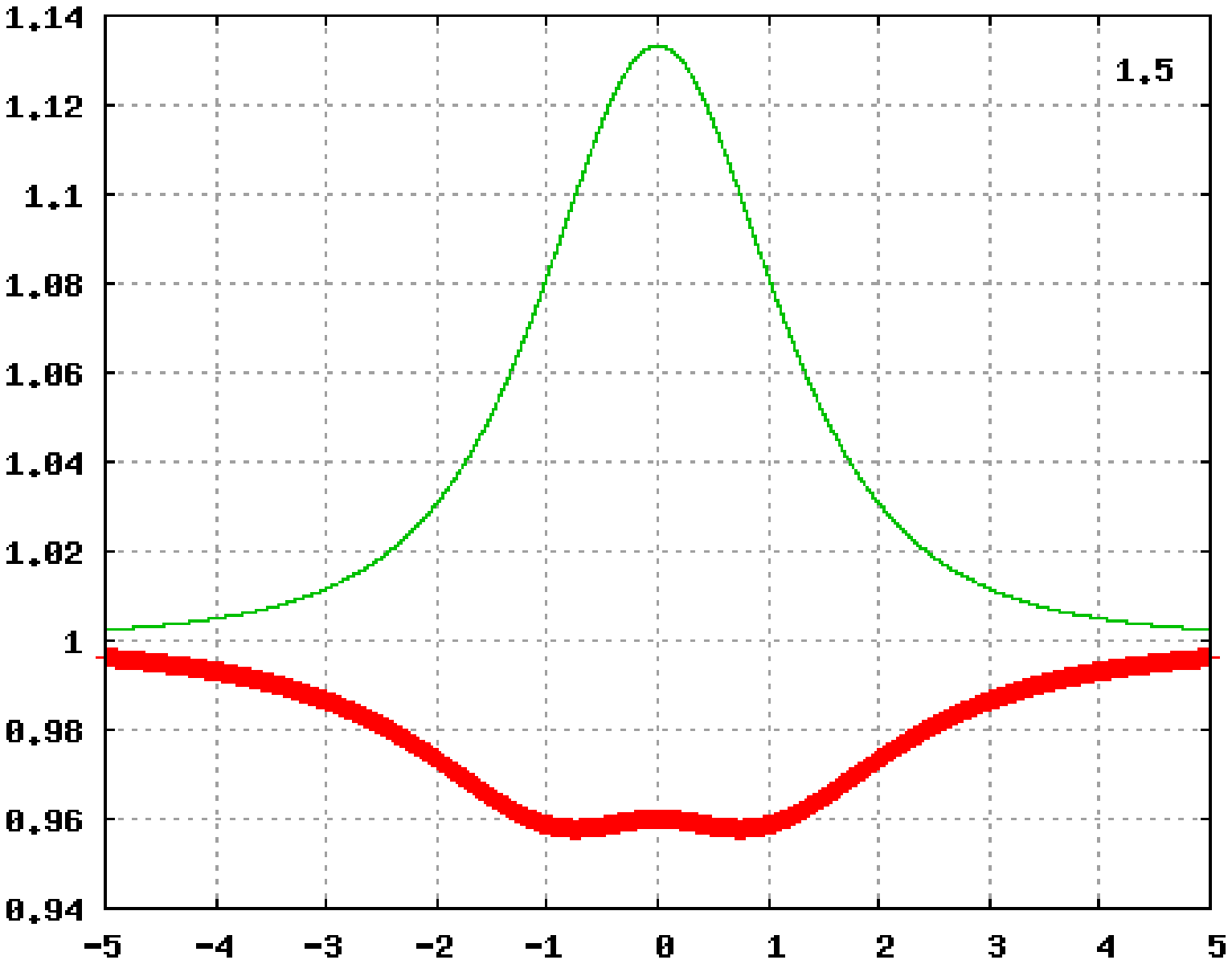}
\caption{Light curves for $\hat{\beta}_0 = 0.2$ (top left), $\hat{\beta}_0 = 0.5 $ (top right), $\hat{\beta}_0 = 1.0$ (bottom left), and $\hat{\beta}_0 = 1.5$ (bottom right). Thick red lines are the light curves for wormholes. Thin green lines are corresponding light curves for Schwarzschild lenses.\label{fig4}}
\end{figure}

The light curve of Schwarzschild lensing was derived by \citet{Pacz86}. The same method of derivation can be used for wormholes. The magnification of the brightness $A$ is
 \begin{eqnarray}
 A  = A_1 + A_2 
      &=& \left|\frac{\hat{\theta}_1}{\hat{\beta}} \frac{d\hat{\theta}_1}{d\hat{\beta}}\right| 
      + \left|\frac{\hat{\theta}_2}{\hat{\beta}} \frac{d\hat{\theta}_2}{d\hat{\beta}}\right|
      =  \left|\frac{\hat{\theta}_1}{\hat{\beta} \left(1 + \frac{2}{\hat{\theta}_1^3}\right)}\right| 
      + \left|\frac{\hat{\theta}_2}{\hat{\beta} \left(1 - \frac{2}{\hat{\theta}_2^3}\right)}\right|, \\ 
      &=& \left|\frac{1}{\left(1 - \frac{1}{\hat{\theta}_1^3}\right) \left(1 + \frac{2}{\hat{\theta}_1^3}\right)}\right| 
               + \left|\frac{1}{\left(1 + \frac{1}{\hat{\theta}_2^3}\right) \left(1 - \frac{2}{\hat{\theta}_2^3}\right)}\right|, 
      \label{eqn:a}
 \end{eqnarray}
 where $A_1$ and $A_2$ are magnification of the outer and inner images,  $\hat{\theta}_1$ and $\hat{\theta}_2$ correspond to outer and inner images, respectively. 
The relation between the lens and source trajectory in the sky is shown in Figure \ref{fig3}. The time dependence of $\hat{\beta}$ is
\begin{equation}
\hat{\beta}(t) = \sqrt{\hat{\beta}_0^2 + {(t -t_0)^2/t_E}^2}, \label{eqn:hbeta}
\end{equation}
where $\hat{\beta}_0$ is the impact parameter of the source trajectory and $t_0$ is the time of closest approach. $t_E$ is the Einstein radius crossing time given by
\begin{equation}
t_E = R_E / v_T,  \label{eqn:te}
\end{equation}
where $v_T$ is the transverse velocity of the lens relative to the source and observer. The light curves obtained from Equations (\ref{eqn:a}) and (\ref{eqn:hbeta}) are shown as thick red lines in Figure \ref{fig4}. The light curves corresponding to Schwarzschild lensing are shown as thin green lines for comparison. The magnifications by the Ellis wormhole are generally less than those of Schwarzschild lensing. The light curve of the Ellis wormhole for $\hat{\beta_o} < 1.0$ shows characteristic gutters on both sides of the peak immediately outside the Einstein ring crossing times ($t = t_0 \pm t_E$). The depth of the gutters is about 4\% from the baseline. Amazingly, the star becomes fainter than normal in terms of apparent brightness in the gutters. This means that the Ellis wormhole lensing has off-center divergence. In conventional gravitational lensing theory \citep{sch92}, the convergence of light is expressed by a convolution of the surface mass density. Thus, we need to introduce negative mass to describe divergent lensing by the Ellis wormhole. However, negative mass is not a physical enitity. As the lensing by the Ellis wormhole is convergent at the center, lensing at some other place must be divergent because the wormhole has zero asymptotic mass. For $\hat{\beta_o} > 1.0$, the light curve of the wormhole has a basin at $t_0$ and no peak. Using these features, discrimination from Schwarzschild lensing can be achieved. Equations (\ref{eqn:re}) and (\ref{eqn:te}) indicate that physical parameters ($D_L$, $a$, and $v_T$) are degenerate in $t_E$ and cannot be derived by fitting the light-curve data. This situation is the same as that for Schwarzschild lensing. To obtain or constrain these values, observations of the finite-source effect \citep{nem94} or parallax \citep{alc95} are necessary.

The detectability of the magnification of the star brightness depends on the timescale. The Einstein radius crossing time $t_E$ depends on the transverse velocity $v_T$. There is no reliable estimate of $v_T$ for wormholes. Here we assume that the velocity of the wormhole is approximately equal to the rotation velocity of stars ($v_T = 220 km/s$) if it is bound to the Galaxy. If the wormhole is not bound to our Galaxy, the transverse velocity would be much higher. We assume $v_T = 5000 km/s$ \citep{saf01} for the unbound wormhole. Table 2 shows the Einstein radius crossing times of the Ellis wormhole lensings for the Galactic bulge and LMC in both bound and unbound scenarios. As the frequencies of current microlensing observations are limited to once every few hours, an event for which the timescale is less than one day is difficult to detect. To find very long timescale events ($t_E \geq 1000 days$), long-term monitoring of events is necessary. The realistic period of observation is $\leq 10 years$. Thus, the realistic size of the throat that we can search for is limited to $100 km \le a \le 10^7 km$ both for the Galactic bulge and LMC if wormholes are bound to our Galaxy. If wormholes are unbound, the detection is limited to $ 10^5 km \le a \le 10^9 km$.

\section{Validity of the weak-field hypothesis}
First, we consider the outer image. In the previous section, we applied the weak-field approximation to the impact parameter $b$ and the deflection angle $\alpha(r)$.
As previously mentioned, the impact parameter $b$ is written as
 \begin{equation}
 b = \sqrt{r^2 + b^2} \approx r (1 + \frac{1}{2} \frac{a^2}{r^2}).
 \end{equation}
The condition to neglect the second term is $a \ll \sqrt{2} r$. As the image is always outside the Einstein ring,
 \begin{equation}
 a \ll \sqrt{2} R_E.
 \end{equation}
From the deflection angle, we obtain a similar relation from Equation (\ref{eqn:defl}): 
 \begin{equation}
 a \ll \sqrt{\frac{8}{5}} R_E.
 \end{equation}
The values of $a$ and $R_E$ in Table \ref{tbl-1} show that the weak-field approximation is suitable for $a \ll 10^{11} km$ in the Galactic microlensing. More generally, $R_E \approx D_S^{1/3} a^{2/3}$ is derived from Equation (\ref{eqn:re}) for $D_L \approx D_S/2$. This means that $R_E$ is much greater than $a$ if $a \ll D_S$. Thus, the weak-field approximation is suitable if the throat radius is negligibly small compared with the source distance. For the inner image, the higher-order effect is expected to be greater than that for the outer image. However, the contribution of the inner image to the total brightness is small and decreases quickly with $\hat{\beta}$ ($A_2/A = 0.034$ for $\hat{\beta} = 2$ and $0.013$ for $\hat{\beta} = 3$). On the other hand, the absolute value of the corresponding $\hat{\theta}$ does not decrease as quickly ($\hat{\theta} = -0.618$ for $\beta = 2$ and $-0.532$ for $\beta = 3$). Thus the contribution of the higher order effect of the second image to the total brightness is expected to be small.  

Another possibility of deviation from the weak-field approximation is the contribution of relativistic images.
Recently, gravitational lensing in the strong-field limit \citep{vir00} has been studied for lensing by black holes. In this limit, light rays are strongly bent and wound close to the photon sphere. As a result, a number of relativistic images appear around the photon sphere. However, it has been shown that there is no photon sphere \citep{dey08} in Ellis wormhole lensing. Therefore, there is no contribution of relativistic images to the magnification in Ellis wormhole lensing. We thus conclude that the weak-field hypothesis is a good approximation unless the throat radius is comparable to the galactic distance.

\section{Optical depth and event rate}
The probability of a microlensing event to occur for a star is expressed by the optical depth $\tau$:
\begin{equation}
\tau = \pi \int_0^{D_S} n(D_L) R_E^2 d D_L,
\end{equation}
where $n(D_L)$ is the number density of wormholes as a function of the line of sight. Here we simply assume that $n(D_L)$ is constant ($n(D_L) = n$):
\begin{eqnarray}
\tau &=& \pi n \int_0^{D_S}  \frac{\pi}{4} \left[ \frac{D_L (D_S - D_L)}{D_S} a^2 \right] ^{2/3}d D_L, \\
       &=& \sqrt[3]{\frac{\pi^5}{2^4}} n a^{4/3} D_S^{5/3} \int_0^1 \left[x ( 1- x)\right]^{2/3} dx, \\
       &\approx& 0.785 n a^{4/3} D_S^{5/3}.
\end{eqnarray}
The event rate expected for a source star $\Gamma$ is calculated as
\begin{eqnarray}
\Gamma &=&  2 \int_0^{D_S} n(D_L) R_E v_T dD_L, \\
                &=& \sqrt[3]{2\pi} n v_T D_S^{4/3} a^{2/3} \int_0^1 \sqrt[3]{x (1 - x)} dx, \\
                &\approx& 0.978 n v_T a^{2/3} D_S^{4/3}. \label{eqn:gm3}
\end{eqnarray}

There is no reliable prediction of the number density of wormholes. Several authors \citep{kra00, lob09} have speculated that wormholes are very common in the universe, at least as abundant as stars. Even if we accept such speculation, there are still large uncertainties in the value of $n$ because the distribution of wormholes is not specified. Here, we introduce two possibilities. One is that wormholes are bound to the Galaxy and the number density is approximately equal to the local stellar density. The other possibility is that wormholes are not bound to the Galaxy and are approximately uniformly distributed throughout the universe. For the bound hypothesis, we use $n = \rho_{Ls}/\langle M_{star} \rangle = 0.147 pc^{-3}$, where $\rho_{Ls}$ is the local stellar density in the solar neighborhood, $\rho_{Ls} = 0.044 M_\odot pc^{-3}$, and $\langle M_{star} \rangle$ is the average mass of stars. We use $\langle M_{star} \rangle = 0.3 M_\odot$, a typical mass of an M dwarf; {\it i.e.}, the dominant stellar component in the Galaxy. For the unbound hypothesis, we assumed that the number density of the wormholes is the same as the average stellar density of the universe. The stellar density of the universe is estimated assuming that the fraction of  baryonic matter accounted for by star is the same as that of the solar neighborhood. Then we obtain $n = \rho_c \Omega_b \rho_{Ls} / (\rho_{Lb} \langle M_{star} \rangle) = 4.97 \times 10^{-9} pc^{-3}$, where $\rho_c = 1.48 \times 10^{-7} M_\odot pc^{-3}$ is the critical density, $\Omega_b = 0.042$ is the baryon density of the universe divided by the critical density, and $\rho_{Ls} = 0.044 M_\odot pc^{-3}$ and $\rho_{Lb} = 0.18 M_\odot pc^{-3}$ are the local star and local baryon densities, respectively.

Using these values, we calculated the optical depths and event rates for bulge and LMC lensings. Table \ref{tbl-3} presents the results for the bulge lensings. In an ordinary Schwarzschild microlensing survey, observations are made of more than 10 million stars. Thus, we can expect approximately $10^7 \Gamma$ events in a year. However, the situation is different in a wormhole search. As mentioned previously, the magnification of wormhole lensing is less than that of Schwarzschild lensing, and a remarkable feature of wormhole lensing is the decreasing brightness around the Einstein radius crossing times. Past microlensing surveys have mainly searched for stars that increase in brightness. The stars monitored are those with magnitudes down to the limiting magnitude or less. However, we need to find stars that decrease in brightness in the wormhole search. To do so, we need to watch brighter stars. Therefore, far fewer stars can be monitored than in an ordinary microlensing survey. Furthermore, the detection efficiency of the wormhole is thought to be less than that for Schwarzschild lensing because of the low magnification. Here we assume that the effective number of stars monitored to find a wormhole is $10^6$. To expect more than one event in a survey of several years, $\Gamma$ must be greater than $\sim 10^{-6}$. The values in Table \ref{tbl-3} indicate that the detection of wormholes with $a > 10^4 km$ is expected in the microlensing survey of the Galactic bulge in the case of the bound model. The results of the optical depths and the event rates for LMC lensing are presented in Table \ref{tbl-4}. On the basis of the same discussion for bulge lensing, we expect $\Gamma > 10^{-6}$ to find a wormhole. The event rates expected for LMC lensing are greater than those for bulge lensing. We expect detection of a wormhole event if $a > 10^2 km$ for the bound model. If no candidate is found, we can set upper limits of $\Gamma$ and/or $\tau$ as functions of $t_E$. To convert these values to physical parameters ($n$ and $a$) requires the distribution of $v_T$. Right now, there is no reliable model of the distribution except for using the bound or unbound hypothesis. On the other hand, the event rates for the unbound model are too small for the events to be detected.

In past microlensing surveys \citep{alc00, tis07, wyr09, sum03}, large amounts of data have already been collected for both the bulge and LMC fields. Monitoring more than $10^6$ stars for about $10 years$ can be achieved by simply reanalyzing the past data. Thus, discovery of wormholes can be expected if their population density is as high as the local stellar density and $10^2 \leqq a \leqq 10^7 km$. Such wormholes of astronomical size are large enough for humans to pass through. Thus, they would be of interest to people discussing the possibility of space--time travel. If no candidate is found, the possibility of a rich population of large-throat wormholes bound to the Galaxy can be ruled out. Such a limit, however, may not affect existing wormhole theories because there is no prediction of the abundance. However, theoretical studies on wormholes are still in progress. The limit imposed by observation is expected to affect future wormhole theories. On the other hand, the discovery of unbound wormholes is very difficult even if their population density is comparable to that of ordinary stars. To discover such wormholes, the monitoring of a much larger number of stars in distant galaxies would be necessary. For example, $\Gamma =  1.7 \times 10^{-6}$ and $t_E \approx 380 days$ for the M101 microlensing survey ($D_S = 7.4 Mpc$) if the throat radius is $10^7 km$. To carry out such a microlensing survey, observation from space is necessary because the resolving of a large number of stars in a distant galaxy is impossible through ground observations. 

Only the Ellis wormhole has been discussed in this paper. There are several other types of wormholes \citep{sh04, nan06, rah07} for which deflection angles have been derived. These wormholes are expected to have different light curves. To detect those wormholes, calculations of their light curves are necessary. The method used in this paper can be employed only when we know the analytic solutions of the image positions. If no analytic solution is found, the calculation must be made numerically.

\section{Summary}
The gravitational lensing of the Ellis wormhole is solved in the weak-field limit. The image positions are calculated as real solutions of simple cubic formulas. One image appears on the source star side and outside the Einstein ring. The other image appears on the other side and inside the Einstein ring. A simple estimation shows that the weak-field hypothesis is a good approximation for Galactic microlensing if the throat radius is less than $10^{11} km$. The light curve derived has characteristic gutters immediately outside the Einstein ring crossing times. Optical depths and event rates for bulge and LMC lensings are calculated for simple bound and unbound hypotheses. The results show that the bound wormholes can be detected by reanalyzing past data if the throat radius is between $10^2$ and $10^7 km$ and the number density is approximately equal to the local stellar density. If the wormholes are unbound and approximately uniformly distributed in the universe with average stellar density, detection of the wormholes is impossible using past microlensing data. To detect unbound wormholes, a microlensing survey of distant galaxies from space is necessary.

\acknowledgments
We would like to thank Professor Hideki Asada of Hirosaki University, Professor Matt Visser of Victoria University, and  Professor Tomohiro Harada of Rikkyo University for discussions and their suggestions concerning this study. We thank Professor Philip Yock of Auckland University for polishing our manuscript. Finally, we would like to thank an anonymous referee who pointed out the existence of a second image.


%% in parentheses, after the keyword, but they will not be verified.

% {\it Facilities:} \facility{Nickel}, \facility{HST (STIS)}, \facility{CXO (ASIS)}.



\appendix

%\clearpage

%% Here we use \plottwo to present two versions of the same figure,
%% one in black and white for print the other in RGB color
%% for online presentation. Note that the caption indicates
%% that a color version of the figure will be available online.
%%

%% This figure uses \includegraphics to scale and rotate the still frame

%% If you are not including electonic art with your submission, you may
%% mark up your captions using the \figcaption command. See the
%% No more than seven \figcaption commands are allowed per page,
%% so if you have more than seven captions, insert a \clearpage

%% Tables should be submitted one per page, so put a \clearpage before

%% Two options are available to the author for producing tables:  the
%% deluxetable environment provided by the AASTeX package or the LaTeX
%% table environment.  Use of deluxetable is preferred.
%%

%% Three table samples follow, two marked up in the deluxetable environment,
%% one marked up as a LaTeX table.

%% In this first example, note that the \tabletypesize{}
%% command has been used to reduce the font size of the table.
%% We also use the \rotate command to rotate the table to
%% landscape orientation since it is very wide even at the
%% Note also that the \label command needs to be placed

%% This table also includes a table comment indicating that the full
%% version will be available in machine-readable format in the electronic

%\clearpage
%\rotate
\begin{center}
\begin{table}
\caption{Einstein radii for bulge and LMC lensings\label{tbl-1}} 
\begin{tabular}{rrrrrrr}
\hline \hline
  & & \multicolumn{2}{c}{Bulge\tablenotemark{a}} & &  \multicolumn{2}{c}{LMC\tablenotemark{b}} \\ \cline{3-4} \cline{6-7}
  $a (km)$ & &  $R_E (km)$ & $\theta_E (mas)$ & & $R_E (km)$ & $\theta_E (mas)$ \\
 \hline
1            & & $3.64 \times 10^5$ & 0.001 & & $6.71 \times 10^5$  & $< 0.001$ \\
10          &  & $1.69 \times 10^6$ & 0.003 & & $3.12 \times 10^6$ & 0.001 \\
$10^2$ &  & $7.85 \times 10^6$ & 0.013 & & $1.45 \times 10^7$ & 0.004 \\
$10^3$ &  & $3.64 \times 10^7$ & 0.061 & & $6.71 \times 10^7$ & 0.018 \\
$10^4$ &  & $1.69 \times 10^8$ & 0.283 & & $3.12 \times 10^8$ & 0.083  \\
$10^5$ &  & $7.85 \times 10^8$ & 1.31    & & $1.45 \times 10^9$ & 0.387  \\
$10^6$ &  & $3.64 \times 10^9$ & 6.10    & &  $6.71 \times 10^9$ &  1.80 \\
$10^7$ &  & $1.69 \times 10^{10}$ & 28.3    & &  $3.12 \times 10^{10}$ &  8.35 \\
$10^8$ &  & $7.85 \times 10^{10}$ & 131    & &  $1.45 \times 10^{11}$ &  38.7 \\
$10^9$ &  & $3.64 \times 10^{11}$ & 610    & &  $6.71 \times 10^{11}$  &  180 \\
$10^{10}$ & & $1.69 \times 10^{12}$ & 2 832    & &  $3.12 \times 10^{12}$ &  835 \\
$10^{11}$ & & $7.85 \times 10^{12}$ & 13 143    & &  $1.45 \times 10^{13}$ &  3 874 \\
%% Text for table notes should follow after the \enddata but before
%% the \end{deluxetable}. Make sure there is at least one \tablenotemark
%% in the table for each \tablenotetext.
\hline \hline
\end{tabular}
\tablecomments{$a$ is the throat radius of the wormhole, $R_E$ is the Einstein radius, and $\theta_E$ is the angular Einstein radius.}
\tablenotetext{a}{$D_S = 8 kpc$ and $D_L = 4 kpc$ are assumed. }
\tablenotetext{b}{$D_S = 50 kpc$ and $D_L = 25 kpc$ are assumed. }
\end{table}
\end{center}

\begin{center}
\begin{table}
\caption{Einstein radius crossing times for bulge and LMC lensings\label{tbl-2}}
\begin{tabular}{rrrrrrrrr}
\hline \hline
  & & \multicolumn{2}{c}{Bulge\tablenotemark{a}} & & \multicolumn{2}{c}{LMC\tablenotemark{b}} \\ 
  $a (km)$ & & \multicolumn{2}{c}{$t_E (day)$} & & \multicolumn{2}{c}{$t_E (day)$}  \\ \cline{3-4} \cline{6-7}
 & & Bound\tablenotemark{c} & Unbound\tablenotemark{d} & & Bound\tablenotemark{c} & Unbound\tablenotemark{d} \\
 \hline
1            & & 0.019 & 0.001 & & 0.035 & 0.002   \\
10          & & 0.089 & 0.004 & & 0.164 & 0.007  \\
$10^2$ & & 0.413 & 0.018 & & 0.761 & 0.033  \\
$10^3$ & & 1.92   & 0.084 & & 3.53   & 0.155  \\
$10^4$ & & 8.90    & 0.392 & &16.4   & 0.721  \\
$10^5$ & & 41.3    & 1.82   & & 76.1   & 3.35  \\
$10^6$ & & 192    & 8.44    & & 353   & 15.5   \\
$10^7$ & & 890    & 39.2  & & 1 639   & 72.1 \\
$10^8$ & & 4 130    & 182    & & 7 608   & 335  \\
$10^9$ & & $> 10^4$    & 843    & & $> 10^4 $  & 1 553  \\
$10^{10}$ & & $> 10^4 $  & 3915    & & $> 10^4$  & 7 212 \\
%% Text for table notes should follow after the \enddata but before
%% the \end{deluxetable}. Make sure there is at least one \tablenotemark
%% in the table for each \tablenotetext.
\hline \hline
\end{tabular}
\tablecomments{$a$ is the throat radius of the wormhole, $t_E$ is the Einstein radius crossing time.}
\tablenotetext{a}{$D_S = 8 kpc$ and $D_L = 4 kpc$ are assumed. }
\tablenotetext{b}{$D_S = 50 kpc$ and $D_L = 25 kpc$ are assumed. }
\tablenotetext{c}{$v_T = 220 km/s$ is assumed.}
\tablenotetext{d}{$v_T = 5000 km/s$ is assumed.}
\end{table}
\end{center}

\begin{center}
\begin{table}
\caption{Optical depths and event rates for bulge lensing\label{tbl-3}} 
\begin{tabular}{rrrrrrrrrrrrrrrrr} 
\hline \hline
 & & \multicolumn{3}{c}{Bound\tablenotemark{a}} & & \multicolumn{3}{c}{Unbound\tablenotemark{b}} \\ \cline{3-5} \cline{7-9}
  $a (km)$ & & \multicolumn{1}{c}{$\tau $} & & $\Gamma (1/year)$ & &   \multicolumn{1}{c}{$\tau $} & & $\Gamma (1/year)$ \\
 \hline
 $10$  & & $8.24 \times 10^{-12}$ & & $2.45 \times 10^{-8}$ & & $2.78 \times 10^{-19} $ & & $1.88 \times 10^{-14}$ \\
 $10^2$  & & $1.77 \times 10^{-10}$ & & $1.14 \times 10^{-7}$ & & $6.00 \times 10^{-18} $ & & $8.73 \times 10^{-14}$ \\
 $10^3$  & & $3.82 \times 10^{-9}$ & & $5.27 \times 10^{-7}$ & & $1.29 \times 10^{-16} $ & & $4.05 \times 10^{-13}$ \\
 $10^4$  & & $8.24 \times 10^{-8}$ & & $2.45 \times 10^{-6}$ & & $2.78 \times 10^{-15} $ & & $1.88 \times 10^{-12}$ \\
 $10^5$  & & $1.77 \times 10^{-6}$ & & $1.14 \times 10^{-5}$ & & $6.00 \times 10^{-14} $ & & $8.73 \times 10^{-12}$ \\
 $10^6$  & & $3.82 \times 10^{-5}$ & & $5.27 \times 10^{-5}$ & & $1.29 \times 10^{-12} $ & & $4.05 \times 10^{-11}$ \\
 $10^7$  & & $8.24 \times 10^{-4}$ & & $2.45 \times 10^{-4}$ & & $2.78 \times 10^{-11} $ & & $1.88 \times 10^{-10}$ \\
 $10^8$  & & $1.77 \times 10^{-2}$ & & $1.14 \times 10^{-3}$ & & $6.00 \times 10^{-10} $ & & $8.73 \times 10^{-10}$ \\
 $10^9$  & & $3.82 \times 10^{-1}$ & & $5.27 \times 10^{-3}$ & & $1.29 \times 10^{-8} $ & & $4.05 \times 10^{-9}$ \\
 $10^{10}$  & & $8.24 $ & & $2.45 \times 10^{-2}$ & & $2.78 \times 10^{-7} $ & & $1.88 \times 10^{-8}$ \\

%% Text for table notes should follow after the \enddata but before
%% the \end{deluxetable}. Make sure there is at least one \tablenotemark
%% in the table for each \tablenotetext.
\hline \hline
\end{tabular}
\tablecomments{$a$ is the throat radius of the wormhole, $\tau$ is the optical depth, $\Gamma$ is the event rate. $D_S = 8 kpc$ is assumed. }
\tablenotetext{a}{$v_T = 220 km/s$ and $n = 0.147 pc^{-3}$ are assumed.}
\tablenotetext{b}{$v_T = 5000 km/s$ and $n = 4.97 \times 10^{-9} pc^{-3}$ are assumed.}
\end{table}
\end{center}

\begin{center}
\begin{table}
\caption{Optical depths and event rates for LMC lensing\label{tbl-4}} 
\begin{tabular}{rrrrrrrrrrrrrrrrr} 
\hline \hline
 & & \multicolumn{3}{c}{Bound\tablenotemark{a}} & & \multicolumn{3}{c}{Unbound\tablenotemark{b}} \\ \cline{3-5} \cline{7-9}
  $a (km)$ & & \multicolumn{1}{c}{$\tau $} & & $\Gamma (1/year)$ & &   \multicolumn{1}{c}{$\tau $} & & $\Gamma (1/year)$ \\
 \hline
 $10$  & & $1.75 \times 10^{-10}$ & & $2.82 \times 10^{-7}$ & & $5.90 \times 10^{-18} $ & & $2.17 \times 10^{-13}$ \\
 $10^2$  & & $3.76 \times 10^{-9}$ & & $1.31 \times 10^{-6}$ & & $1.27 \times 10^{-16} $ & & $1.01 \times 10^{-12}$ \\
 $10^3$  & & $8.11 \times 10^{-8}$ & & $6.07 \times 10^{-6}$ & & $2.74 \times 10^{-15} $ & & $4.67 \times 10^{-12}$ \\
 $10^4$  & & $1.75 \times 10^{-6}$ & & $2.82 \times 10^{-5}$ & & $5.90 \times 10^{-14} $ & & $2.17 \times 10^{-11}$ \\
 $10^5$  & & $3.76 \times 10^{-5}$ & & $1.31 \times 10^{-4}$ & & $1.27 \times 10^{-12} $ & & $1.01 \times 10^{-10}$ \\
 $10^6$  & & $8.11 \times 10^{-4}$ & & $6.07 \times 10^{-4}$ & & $2.74 \times 10^{-11} $ & & $4.67 \times 10^{-10}$ \\
 $10^7$  & & $1.75 \times 10^{-2}$ & & $2.82 \times 10^{-3}$ & & $5.90 \times 10^{-10} $ & & $2.17 \times 10^{-9}$ \\
 $10^8$  & & $3.76 \times 10^{-1}$ & & $1.31 \times 10^{-2}$ & & $1.27 \times 10^{-8} $ & & $1.01 \times 10^{-8}$ \\
 $10^9$  & & $8.11 $ & & $6.07 \times 10^{-2}$ & & $2.74 \times 10^{-7} $ & & $4.67 \times 10^{-8}$ \\
 $10^{10}$  & & $175 $ & & $2.82 \times 10^{-1}$ & & $5.90 \times 10^{-6} $ & & $2.17 \times 10^{-7}$ \\

%% Text for table notes should follow after the \enddata but before
%% the \end{deluxetable}. Make sure there is at least one \tablenotemark
%% in the table for each \tablenotetext.
\hline \hline
\end{tabular}
\tablecomments{$a$ is the throat radius of the wormhole, $\tau$ is the optical depth, $\Gamma$ is the event rate. $D_S = 8 kpc$ is assumed. }
\tablenotetext{a}{$v_T = 220 km/s$ and $n = 0.147 pc^{-3}$ are assumed.}
\tablenotetext{b}{$v_T = 5000 km/s$ and $n = 4.97 \times 10^{-9} pc^{-3}$ are assumed.}
\end{table}
\end{center}

%% If you use the table environment, please indicate horizontal rules using
%% Do not put multiple tabular environments within a single table.
%% The optional \label should appear inside the \caption command.

%% If the table is more than one page long, the width of the table can vary
%% from page to page when the default \tablewidth is used, as below.  The
%% individual table widths for each page will be written to the log file; a
%% maximum tablewidth for the table can be computed from these values.
%% The \tablewidth argument can then be reset and the file reprocessed, so
%% that the table is of uniform width throughout. Try getting the widths
%% from the log file and changing the \tablewidth parameter to see how
%% adjusting this value affects table formatting.

%% The \dataset{} macro has also been applied to a few of the objects to
%% show how many observations can be tagged in a table.

%% Tables may also be prepared as separate files. See the accompanying
%% sample file table.tex for an example of an external table file.
%% To include an external file in your main document, use the \input
%% command. Uncomment the line below to include table.tex in this
%% sample file. (Note that you will need to comment out the \documentclass,
%% \begin{document}, and \end{document} commands from table.tex if you want
%% to include it in this document.)


%% The following command ends your manuscript. LaTeX will ignore any text

\end{document}